\documentclass[twocolumn,showpacs,preprintnumbers,amsmath,amssymb]{revtex4}

\usepackage{graphicx}
\usepackage{dcolumn}
\usepackage{bm}
\usepackage{psfig}
\usepackage{times}
\newcommand{\be}{\beta}

\def\be{\begin{equation}}
\def\ee{\end{equation}}

\begin{document}


\title{Thin-shell wormholes from regular charged black holes}

\author{F. Rahaman}
 \email{farook\_rahaman@yahoo.com}
\affiliation{Department of Mathematics, Jadavpur University,
Kolkata 700 032, West Bengal, India}

\author{Peter K. F. Kuhfittig }
\email{kuhfitti@msoe.edu} \affiliation{Department of Mathematics,
Milwaukee School of Engineering, Milwaukee, Wisconsin 53202-3109,
USA}

\author{K. A. Rahman}
  \email{farook\_rahaman@yahoo.com}\affiliation{Department of Mathematics, Jadavpur
University, Kolkata 700 032, West Bengal, India}

\author{Sk. A. Rakib}
 \email{farook\_rahaman@yahoo.com} \affiliation{Department of Mathematics, Jadavpur
University, Kolkata 700 032, West Bengal, India}

\date{\today}

\begin{abstract}\noindent
We investigate  a new thin-shell wormhole constructed by
surgically grafting  two regular charged black holes arising
from  the action using  nonlinear electrodynamics coupled to
general relativity. The stress-energy components within the
shell violate the null and weak energy conditions but
obey the strong energy condition. We study the stability in
two ways: (i) taking a specific equation of state at the throat
and (ii) analyzing the stability to linearized spherically
symmetric perturbations about a static equilibrium solution.
Various other aspects of this thin-shell wormhole are also
analyzed.
\end{abstract}


\maketitle

\section{Introduction}\noindent
Over 20 years ago Visser \cite{Visser1989} proposed a theoretical
method for constructing a new class of wormholes from a black-hole
spacetime. This type of wormhole is known as a thin-shell wormhole
and is constructed by applying the so-called cut-and-paste
technique: surgically graft two black-hole spacetimes  together in
such a way that no event horizon is permitted to form. This method
yields a wormhole spacetime whose throat is a time-like
hypersurface, i.e., a three-dimensional thin shell. Since Visser's
novel approach yields a way of minimizing the use of exotic matter
to construct a wormhole, the technique was quickly adopted by
various authors for constructing thin-shell wormholes
\cite{Poisson1995,Lobo2003,Lobo2004,Eiroa2004a,
Eiroa2004b,Eiroa2005,Thibeault2005,Lobo2005,Rahaman2006,
Eiroa2007,Rahaman2007a,Rahaman2007b,Rahaman2007c,Richarte2007,Lemos2008,Rahaman2008a,Rahaman2008b,Eiroa2008a,Eiroa2008b}.

In 1999,  E. A. Beato and A. Garcia \cite{Beato1999}
discovered a new regular exact black hole solution which
comes from the action using nonlinear electrodynamics
coupled to general relativity.  The dynamics of the
theory is governed by the action
\begin{equation}
I = \frac{1}{16 \pi } \int  d^4x \sqrt{-g}\left[ R - L(F)\right].
\end{equation}
Here the nonlinear electrodynamics is described by a type of
gauge-invariant Lagrangian $L(F)$ , where $F_{\mu\nu}$ is the
Maxwell field tensor, $F$ is the contracted Maxwell scalar, i.e.,
$F^\mu_\mu=F$, while $R$ is the curvature scalar.

To obtain the desired solution  from the  above action (1),
Beato and Garcia considered a static and spherically
symmetric configuration given by

\begin{equation}
ds^2 = -f(r) dt^2 + f(r)^{-1}dr^2 + r^2 (d\theta^2+\sin^2\theta
d\phi^2),
\end{equation}

where
\begin{equation}
f(r) = 1- \frac{2M}{r} + \frac{2M}{r}\tanh
\left(\frac{Q^2}{2Mr}\right).
\end{equation}
Here the parameters $M$ and $Q$ can be  associated with mass and
charge, respectively, of the black hole.

It is shown in Ref. \cite{Beato1999} that this black hole has
two event horizons $r_-$ and $r_+$ whenever $|Q|\le 1.05M$.  So
for suitable choices of the parameters $M$ and $Q$, the points
$r_-$ and $r_+$ are simply the $r$-intercepts of $f(r)$.
(See Figure 1.)
\begin{figure}
\begin{center}
\vspace{0.5cm}
\includegraphics[width=0.3\textwidth]{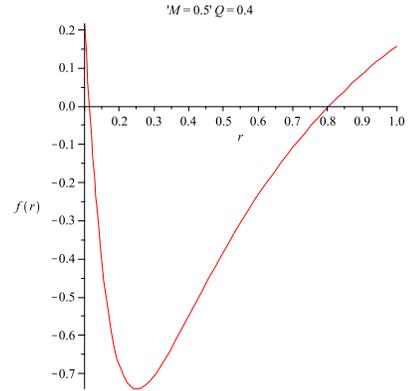}
\caption{Event horizons occur at $r_-$ and $r_+$, where $f(r)$ cuts
the $r$-axis, using suitable parameters.  Here $ M = 0.5$ and
  $ Q =0.4 $.}
\end{center}
\end{figure}

The purpose of this paper is to employ this class of regular
charged black holes to construct a traversable thin-shell
wormhole by means of the cut-and-paste technique.

\section{Thin-Shell Wormhole Construction}\noindent
The mathematical construction of our thin-shell wormhole begins
by taking two copies of the regular charged black hole and
removing from each the four-dimensional region
\[
  \Omega^\pm = \{r\leq a\,|\,a>r_h\}.
\]
Here $r_h=r_+$, the larger of the two radii.  We now identify
(in the sense of topology) the timelike hypersurfaces
\[
  \partial\Omega^\pm =\{r=a\,|\,a>r_h\},
\]
denoted by $\Sigma$.  The resulting manifold is geodesically
complete and consists of two asymptotically flat regions
connected by a throat. The induced metric on $\Sigma$ is
given by
\begin{equation}
               ds^2 =  - d\tau^2 + a(\tau)^2( d\theta^2 +
               \sin^2\theta d\phi^2),
\end{equation}
where $\tau$ is the proper time on the junction surface.  Using
the Lanczos equations
\cite{Visser1989,Poisson1995,Lobo2003,Lobo2004,Eiroa2004a,Eiroa2004b,Eiroa2005,
Thibeault2005,Lobo2005,Rahaman2006,
Eiroa2007,Rahaman2007a,Rahaman2007b,Rahaman2007c,Richarte2007,Lemos2008,Rahaman2008a,
Rahaman2008b,Eiroa2008a,Eiroa2008b},
\begin{equation*}\label{E:Lanczos}
  S^i_{\phantom{i}j}=-\frac{1}{8\pi}\left([K^i_{\phantom{i}j}]
   -\delta^i_{\phantom{i}j}[K]\right),
\end{equation*}
one can obtain the surface stress-energy tensor
$S^i_{\phantom{i}j}=\text{diag}(-\sigma, p_{\theta},
 p_{\phi})$, where $\sigma$ is the surface-energy density and
$p=p_{\theta}=p_{\phi}$ is the surface pressure. The Lanczos
equations now yield
\begin{equation*}\label{E:stress1}
  \sigma=-\frac{1}{4\pi}[K^\theta_{\phantom{\theta}\theta}]
\end{equation*}
and
\begin{equation*}\label{E:stress2}
  p=\frac{1}{8\pi}\left([K^\tau_{\phantom{\tau}\tau}]
    +[K^\theta_{\phantom{\theta}\theta}]\right).
\end{equation*}

A dynamic analysis can be obtained by letting the radius $r=a$
be a function of time \cite{Poisson1995}.  As a result,
\begin{equation}\label{E:sigma}
\sigma = - \frac{1}{2\pi a}\sqrt{f(a) + \dot{a}^2}
\end{equation}
and
\begin{equation}\label{E:pressure}
p_{\theta} = p_{\phi} = p =  -\frac{1}{2}\sigma + \frac{1}{8\pi
}\frac{2\ddot{a} + f^\prime(a) }{\sqrt{f(a) + \dot{a}^2}}.
\end{equation}
Here $ p$ and $\sigma $ obey the conservation equation

\begin{equation}
               \frac {d}{d \tau} (\sigma a^2) + p \frac{d}{d \tau}(a^2)= 0
               \end{equation}
or
\begin{equation}\label{E:conservation}
               \dot{\sigma} + 2 \frac{\dot{a}}{a}( p + \sigma ) = 0.
               \end{equation}
In the above equations, the overdot and prime denote, respectively,
the derivatives with respect to $\tau$ and $a$.

For a static configuration of radius $a$, we obtain (assuming
$\dot{a} = 0 $ and $\ddot{a}= 0 $) from  Eqs. (5) and (6)
\begin{equation}
\sigma = - \frac{1}{2 \pi a }\left[1- \frac{2M}{a} +
\frac{2M}{a}\tanh
\left(\frac{Q^2}{2Ma}\right)\right]^{\frac{1}{2}}
\end{equation}
and
\begin{equation}
p  = \frac{\left[1- \frac{M}{a} + \frac{M}{a}\tanh
\left(\frac{Q^2}{2Ma}\right)-\frac{Q^2}{2a^2}\cosh^{-2}
\left(\frac{Q^2}{2Ma}\right)\right]}{4 \pi a \left[1-
\frac{2M}{a} + \frac{2M}{a}\tanh
\left(\frac{Q^2}{2Ma}\right)\right]^{\frac{1}{2}}}.
\end{equation}

Observe that the energy-density $\sigma$, as well as $\sigma+p$,
are negative. So the shell contains matter that violates both the
null energy condition (NEC) and the weak energy condition (WEC).
Also, since $ \sigma + 2p $  $ $ and $\sigma+3p $ are positive,
the strong energy condition is satisfied. \\Using various values
of the parameters $M$ and $Q$, Figures 2-9 show the plots for
$\sigma$ and $p$ as functions of the radius $a$. We choose
typical wormholes whose radii  fall within the range 0.01 to 10
km.

\section{Equation of state}

\begin{figure}
\begin{center}
\vspace{0.5cm}
\includegraphics[width=0.3\textwidth]{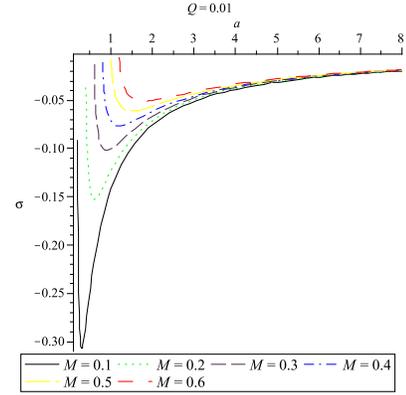}
\caption{Plots for $\sigma$ versus $a$. We choose the fixed
value $Q=0.01$ and various values for $M$.} \label{fig3}
\end{center}
\end{figure}

\begin{figure}
\begin{center}
\vspace{0.5cm}
\includegraphics[width=0.3\textwidth]{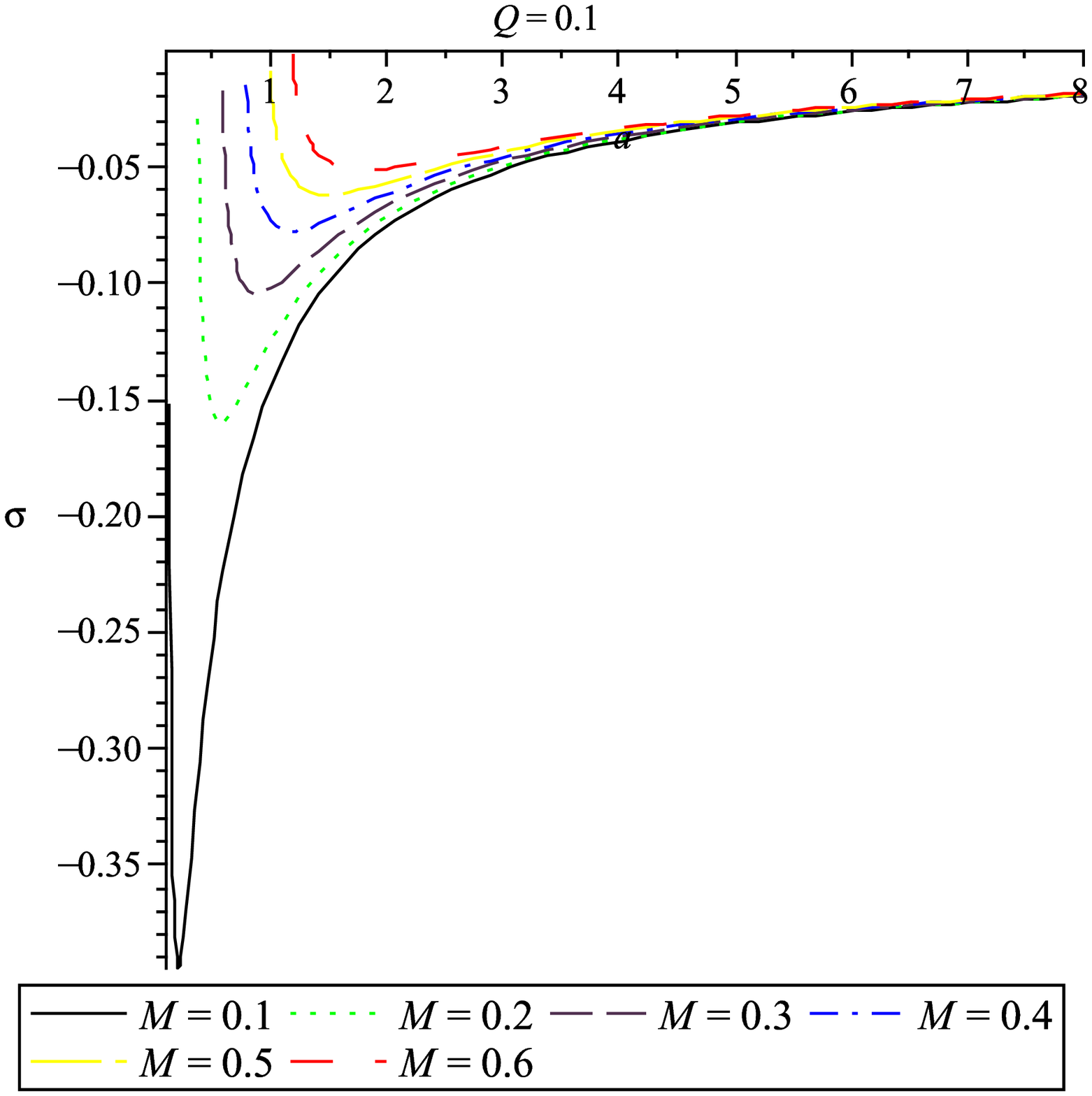}
\caption{Plots for $\sigma$ versus $a$. We choose the fixed
value $Q=0.1$ and various values for $M$.} \label{fig4}
\end{center}
\end{figure}

\begin{figure}
\begin{center}
\vspace{0.5cm}
\includegraphics[width=0.3\textwidth]{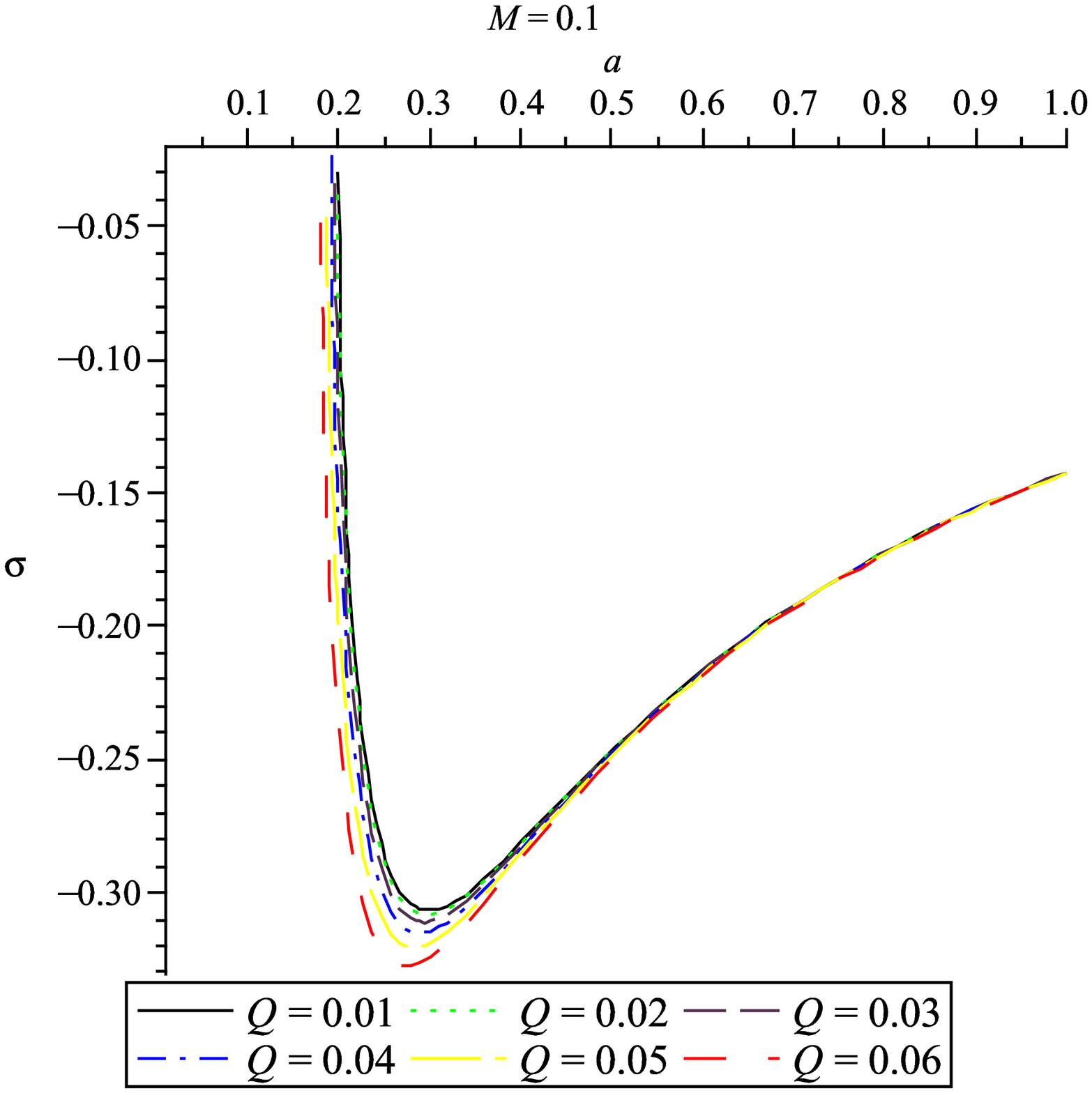}
\caption{Plots for $\sigma$ versus $a$. We choose the fixed
value $M=0.1$ and various values for $Q$.} \label{fig6}
\end{center}
\end{figure}

\begin{figure}
\begin{center}
\vspace{0.5cm}
\includegraphics[width=0.3\textwidth]{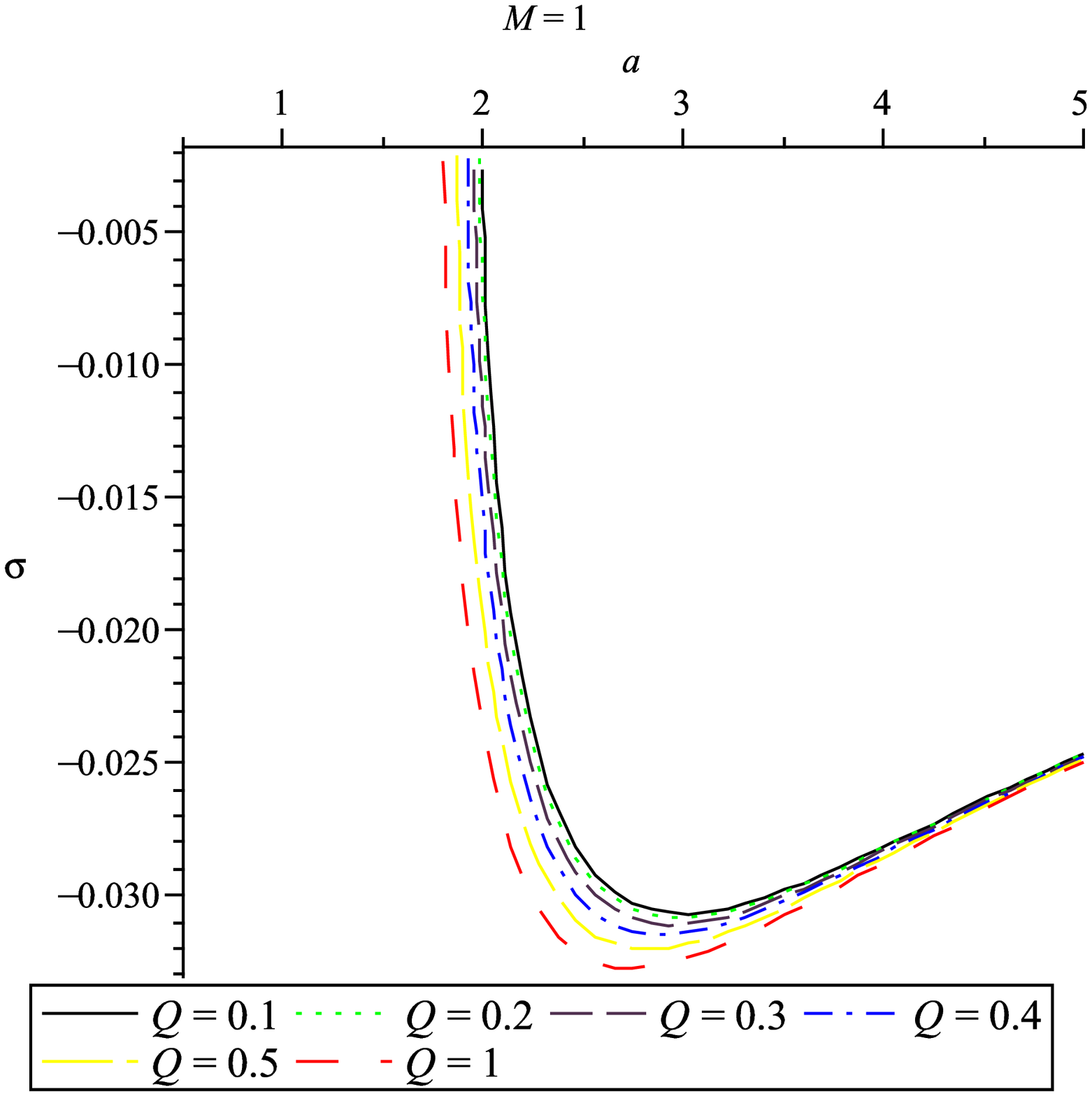}
\caption{Plots for $\sigma$ versus $a$. We choose the fixed
 value $M=1$ and various values for $Q$.} \label{fig8}
\end{center}
\end{figure}

\begin{figure}
\begin{center}
\vspace{0.5cm}
\includegraphics[width=0.3\textwidth]{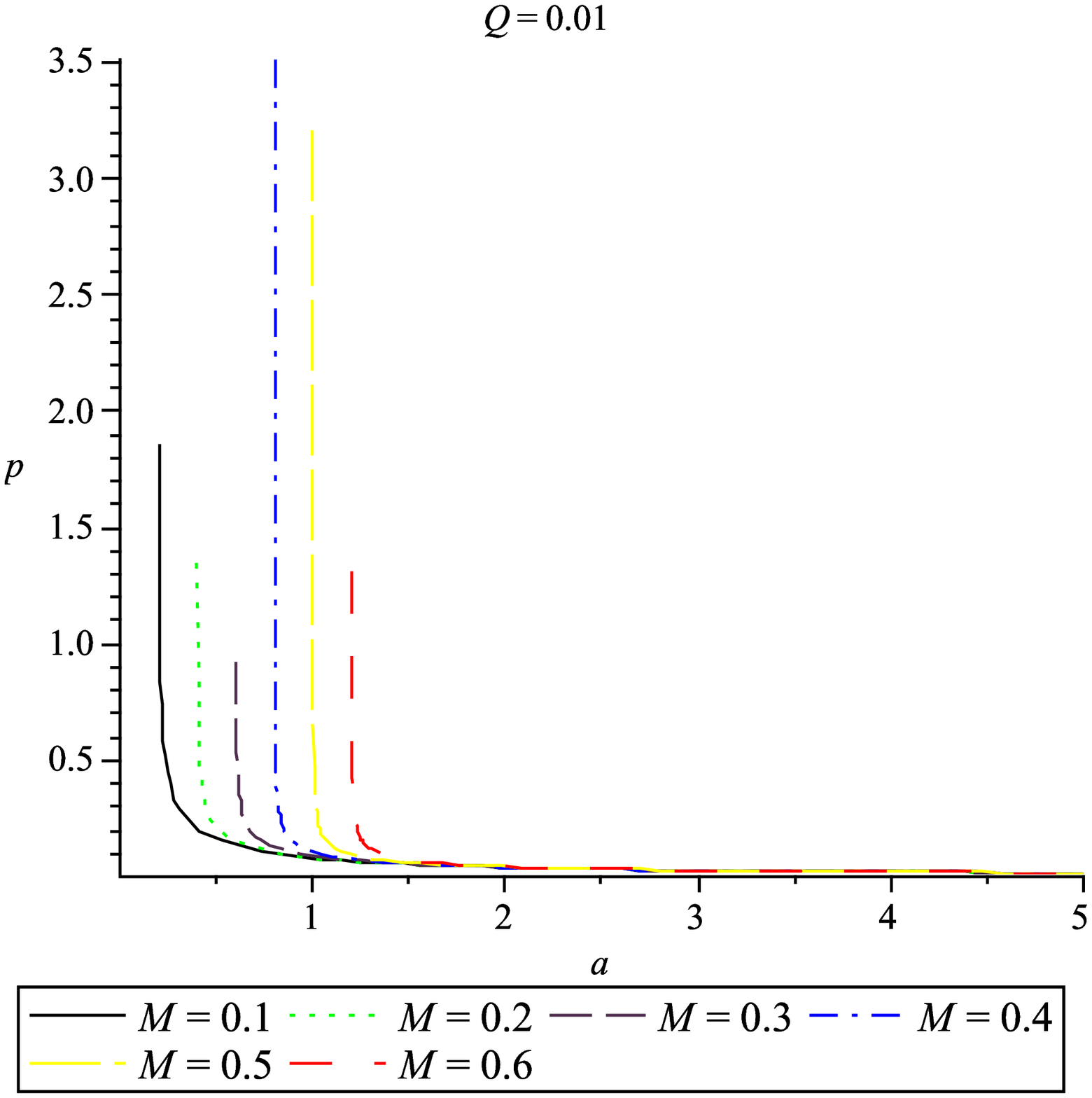}
\caption{Plots for $p$ versus $a$. We choose the fixed value
 $Q=0.01$ and various values for $M$.} \label{fig9}
\end{center}
\end{figure}

\begin{figure}
\begin{center}
\vspace{0.5cm}
\includegraphics[width=0.3\textwidth]{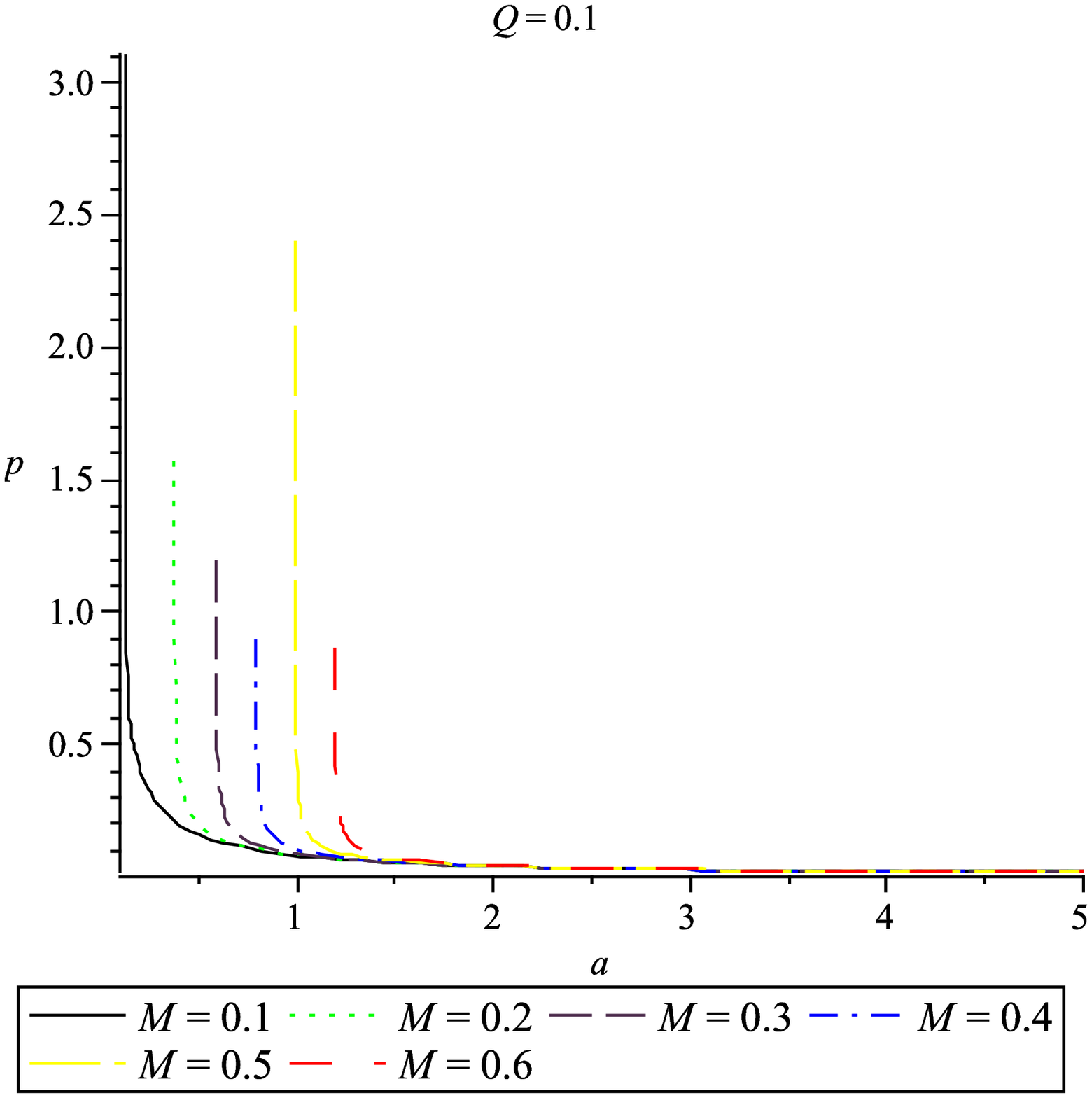}
\caption{Plots for $p$ versus $a$. We choose the fixed value
$Q=0.1$ and various values for $M$.} \label{fig10}
\end{center}
\end{figure}

\begin{figure}
\begin{center}
\vspace{0.5cm}
\includegraphics[width=0.3\textwidth]{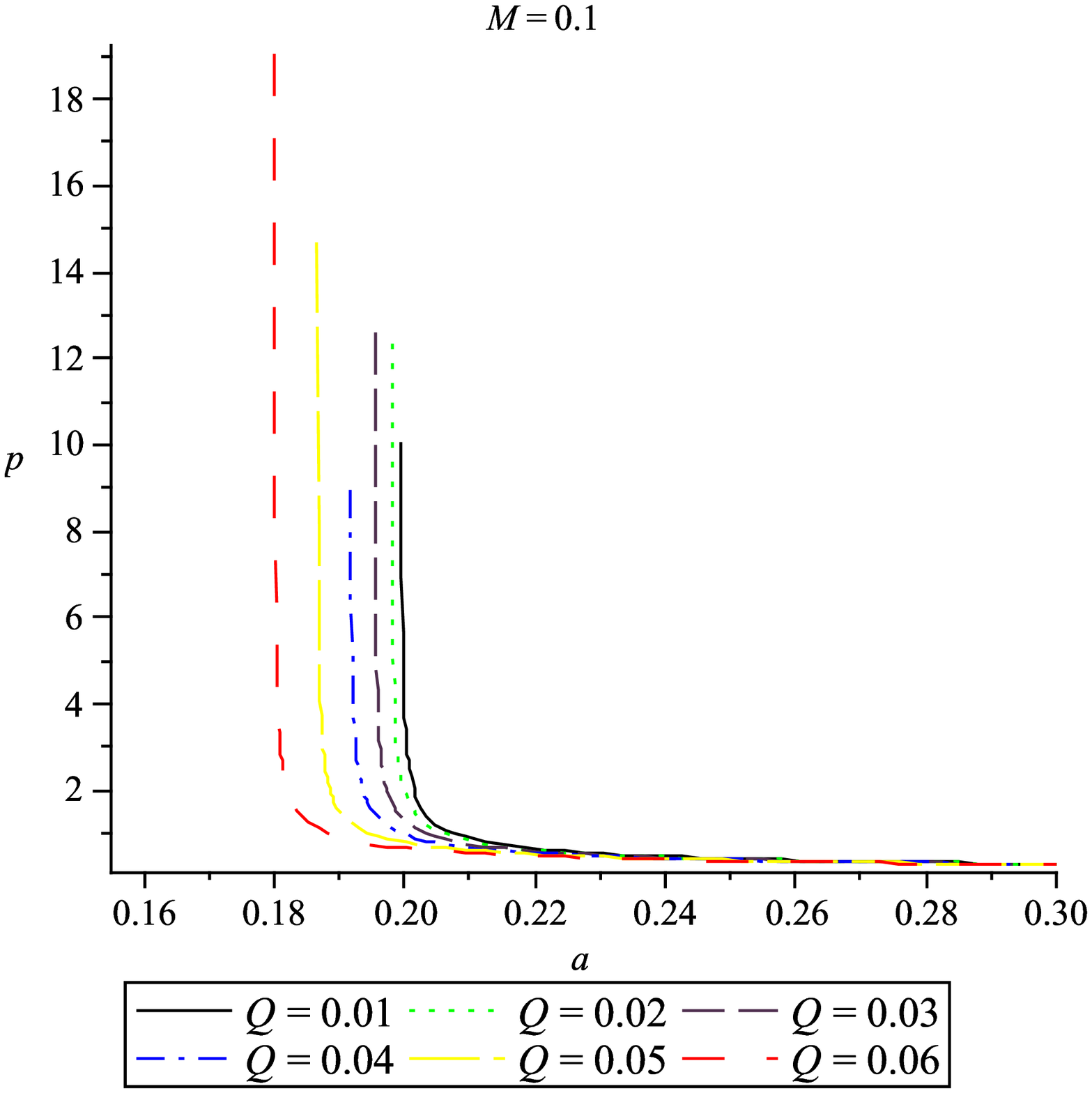}
\caption{Plots for $p$ versus $a$. We choose the fixed
value $M=0.1$ and various values for $Q$.} \label{fig12}
\end{center}
\end{figure}

\begin{figure}
\begin{center}
\vspace{0.5cm}
\includegraphics[width=0.3\textwidth]{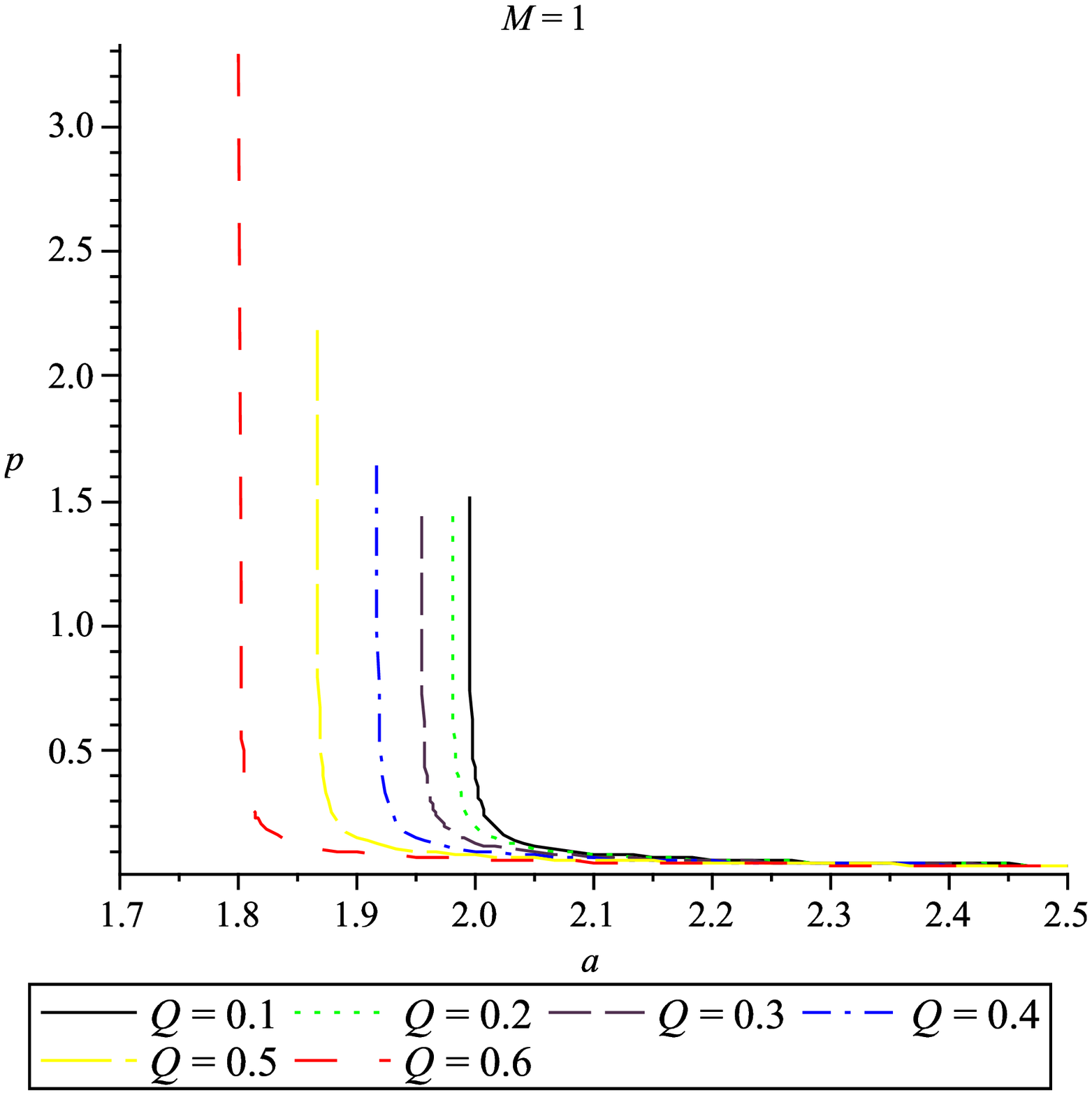}
\caption{Plots for $p$ versus $a$. We choose the fixed value
$M=1$ and various values for $Q$.} \label{fig14}
\end{center}
\end{figure}

\noindent Let us suppose that the EoS at the surface
$\Sigma$ is $p=w\sigma, w\equiv\text{constant}$. From
Eqs. (9) and (10),
\begin{multline}
\frac{p}{\sigma}  = w = \\-\frac{1}{2}-  \frac{M}{2a}
\frac{\left[1- \tanh
\left(\frac{Q^2}{2Ma}\right)-\frac{Q^2}{2Ma}\cosh^{-2}
\left(\frac{Q^2}{2Ma}\right)\right]}{4 \pi a \left[1-
\frac{2M}{a} + \frac{2M}{a}\tanh
\left(\frac{Q^2}{2Ma}\right)\right]}.
\end{multline}
Observe that if $a \rightarrow \infty$, i.e., the location of the
wormhole throat is large enough, then $w\rightarrow -\frac{1}{2}$.
When $a\rightarrow a_0 $, where $a_0$ is the point where the curve
cuts the $a$-axis in Fig 10, then $p \rightarrow 0$, which
would normally be viewed as a dust shell.  Since $a_0 < r_h $,
however, the dust shell is never be found. On the other hand, since
the Casimir effect with a massless field is of the traceless type,
it may be of interest to check the traceless surface stress-energy
tensor, $S^i_{\phantom{i}j}=0$, i.e., $ -\sigma + 2p =0 $.  From
this equation we find that

\begin{figure}
\begin{center}
\vspace{0.5cm}
\includegraphics[width=0.3\textwidth]{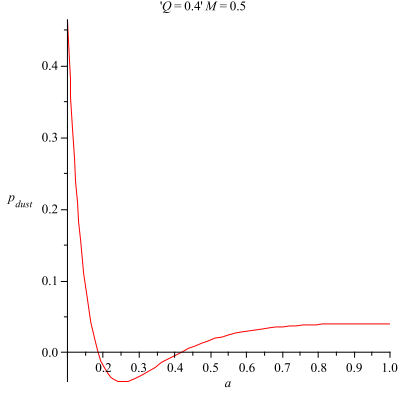}
\caption{The curve cuts the $a$-axis at $a_0 < r_h$; for the given
plot, $ M = 0.5$ and $ Q =0.4. $ }. \label{fig17}
\end{center}
\end{figure}

\begin{multline}
  g(a) \equiv 4- \frac{6M}{a} + \frac{6M}{a}
\tanh \left(\frac{Q^2}{2Ma}\right)\\ -\frac{Q^2}{a^2}\cosh^{-2}
\left(\frac{Q^2}{2Ma}\right)  = 0.
\end{multline}
Figure 11 indicates that the value of $a$ satisfying this
equation is inside the event horizon ($r=r_h$)  of the regular
black hole.  It follows that this situation cannot arise in a
wormhole setting.

\begin{figure}
\begin{center}
\vspace{0.5cm}
\includegraphics[width=0.3\textwidth]{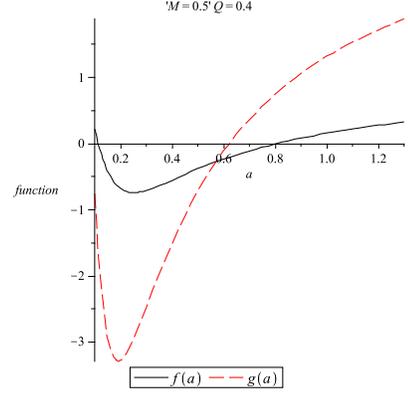}
\caption{The curve $g(a)$  cuts the $a$-axis at a point less
than $r_h$; for the given plot, $ M = 0.5$ and $ Q =0.4 $ .  }
\label{fig17}
\end{center}
\end{figure}

\section{The Gravitational field}
\noindent
In this section we are going to take a brief look at
the attractive or repulsive nature of our wormhole.  To
do so, we calculate the observer's four-acceleration
\[ a^\mu = u^\mu_{; \nu} u^\nu, \]
where
\[u^{\nu} = \frac {d x^{\nu}}{d {\tau}} = \left( \frac{1}{\sqrt{f(r)}},
0,0,0\right).\] Taking into account Eq. (2), the only nonzero
component is given by
\[ a^r = \Gamma^r_{tt} \left(\frac{dt}{d\tau}\right)^2
= \frac{M}{r^2} \alpha(r), \] where
 \begin{equation}
 \alpha(r) = \left  [1- \tanh \left(\frac{Q^2}{2Mr}\right)
-\frac{Q^2}{2Mr}\cosh^{-2} \left(\frac{Q^2}{2Mr}\right)\right].
  \end{equation}
A test particle   moving radially from rest obeys the geodesic equation
\[ \frac{d^2r}{d\tau^2}= -\Gamma^r_{tt}\left(\frac{dt}{d\tau}\right)^2 = -a^r.
\]
It is true in general that a wormhole is attractive whenever $
a^r>0 $.  In our situation, $ a^r$ is positive for $ r >
\frac{Q^2}{2Mr_0} $, where $r_0$ is the point where $ \alpha(r) $
cuts the $x$-axis in Fig. 12. In other words, the wormhole is
attractive for $ r >\frac{Q^2}{2Mr_0}$  and repulsive for $ r
<\frac{Q^2}{2Mr_0}$. Finally, an observer at rest is a geodesic
observer whenever $ r = \frac{Q^2}{2Mr_0}$.

 \begin{figure}
\begin{center}
\vspace{0.5cm}
\includegraphics[width=0.3\textwidth]{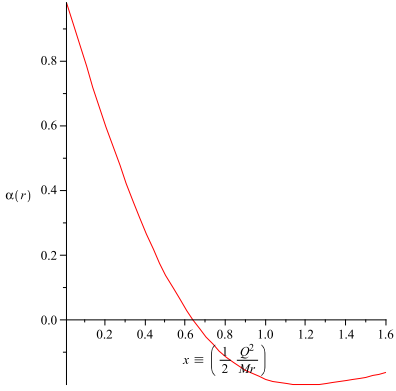}
\caption{ $r_0$ is the point where $ \alpha (r)$ cuts the $x$-axis.
} \label{fig17}
\end{center}
\end{figure}

\section{The Total amount of exotic matter}
\noindent
In this section we determine the total amount of exotic matter
for the thin-shell wormhole.  This total can be quantified by the
integral
 \cite{Eiroa2005,Thibeault2005,Lobo2005,Rahaman2006,
Eiroa2007,Rahaman2007a,Rahaman2007b}
\begin{equation}
   \Omega_{\sigma}=\int [\rho+p]\sqrt{-g}d^3x.
\end{equation}
By introducing the radial coordinate $R=r-a$, w get
\[
 \Omega_{\sigma}=\int^{2\pi}_0\int^{\pi}_0\int^{\infty}_{-\infty}
     [\rho+p]\sqrt{-g}\,dR\,d\theta\,d\phi.
\]
Since the shell is infinitely thin, it does not exert any radial
pressure.  Moreover, $\rho=\delta(R)\sigma(a)$.  So
\begin{multline}
 \Omega_{\sigma}=\int^{2\pi}_0\int^{\pi}_0\left.[\rho\sqrt{-g}]
   \right|_{r=a}d\theta\,d\phi=4\pi a^2\sigma(a)\\
   =4\pi a^2\left(-\frac{1}{2\pi a}\right)
           \sqrt{1-\frac{2M}{a}+\frac{2M}{a}\text{tanh}
                \left(\frac{Q^2}{2Ma}\right)}\\
  =-2a\sqrt{1-\frac{2M}{a}+\frac{2M}{a}\text{tanh}
                \left(\frac{Q^2}{2Ma}\right)}.
\end{multline}

 This NEC violating matter ($\Omega_\sigma $) can be reduced by
choosing a value for $a$ closer to $r=r_h$.  The closer $a$ is
to $r_h$, however, the closer the wormhole is to a black hole:
incoming microwave background radiation would get blueshifted
to an extremely high temperature \cite{tR93}.  On the other
hand, it follows from Eq. (15) that for $a\gg r_h$,
$\Omega_{\sigma}$ will  depend linearly on $a$: \\
\begin{equation}
\Omega_{\sigma} \approx -2a. \end{equation} The variation of the
total amount of exotic matter with respect to the mass and charge
of the black hole can best be seen graphically (Figures 13 - 15).
Observe that the mass on the thin shell can be reduced by either
increasing the  mass or decreasing  the charge of the black hole.

\begin{figure}
\begin{center}
\vspace{0.5cm}
\includegraphics[width=0.3\textwidth]{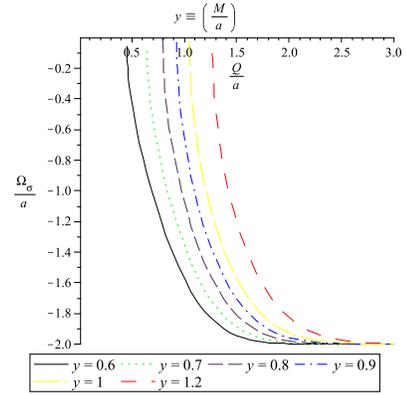}
\caption{The variation in the total amount of exotic matter on the
shell with respect to the charge of the black hole, while assuming
a fixed mass for the black hole.}
        \label{fig17}
\end{center}
\end{figure}

\begin{figure}
\begin{center}
\vspace{0.5cm}
\includegraphics[width=0.3\textwidth]{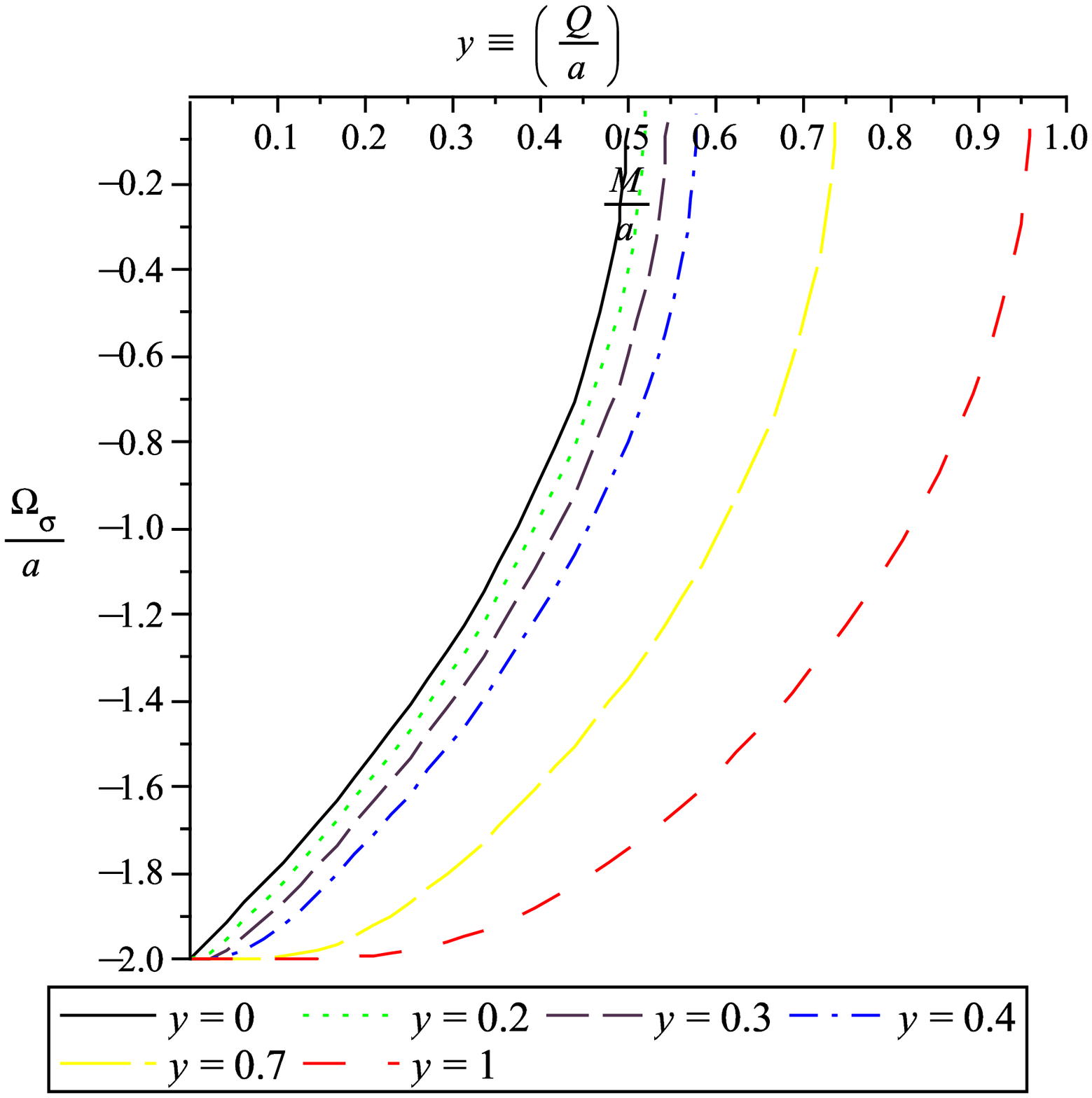}
\caption{The variation in the total amount of exotic matter on the
shell with respect to the mass of the black hole, while assuming
 a fixed charge for the black hole.}
        \label{fig17}
\end{center}
\end{figure}

\begin{figure}
\begin{center}
\vspace{0.5cm}
\includegraphics[width=0.4\textwidth]{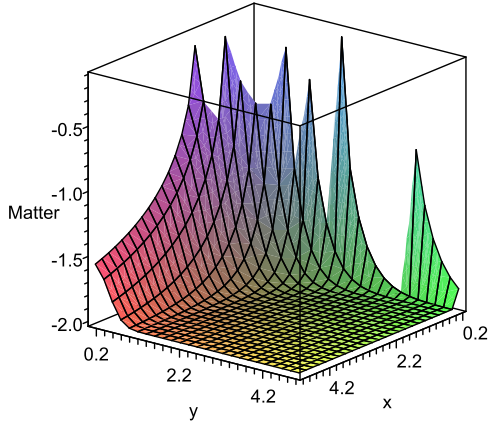}
\caption{The variation in the total amount of exotic matter on the
shell with respect to the mass ($ x= \frac{M}{a} $) and the charge
($ y= \frac{Q}{a} $) of the black hole.}
        \label{fig17}
\end{center}
\end{figure}

\pagebreak

\section{Stability}
\noindent
In this section we turn to the question of stability of the
wormhole using two different approaches: (i) assuming a
specific equation of state on the thin shell and (ii)
analyzing the stability to linearized radial perturbations.

 \subsection{Reintroducing the equation of state} \noindent
Suppose we return to the EoS
\begin{equation}
   p=w\sigma, \quad w<0,
\end{equation}
which is analogous to the equation of state normally
associated with dark energy.  Then Eq. (\ref{E:conservation})
yields
\begin{equation} \sigma(a) = \sigma_0 \left( \frac{a_0}{a} \right)^{2(1+w)},    \end{equation}
where $a=a_0$ is the static solution and $ \sigma_0
= \sigma(a_0) $.
Rearranging Eq. (\ref{E:sigma}), we obtain the thin
shell's equation of motion
\begin{equation}
\dot{a}^2 + V(a)= 0.
\end{equation}
Here the potential $V(a)$ is defined  as
\begin{equation}\label{E:potential}
V(a) =  f(a) - \left[2\pi a \sigma(a)\right]^2.
\end{equation}
Substituting the value of $\sigma(a)$ in this equation,
we obtain the following form of the potential:
\begin{equation}
V(a) = 1- \frac{2M}{a} + \frac{2M}{a}\tanh
\left(\frac{Q^2}{2Ma}\right) - \frac{ A }{a^{2+4w}},
\end{equation}
where $ A = 4 \pi^2 \sigma_0^2a_0^{4+4w}$.

If $-1<w<0$, then $p+\sigma=\sigma w+\sigma=\sigma
(1+w)<0$, so that the WEC is violated.  Figure 16 shows
that the wormhole is stable for a certain range of
parameters.  (We will examine this range more closely
in the next subsection.)

If $w<-1$, then $p+\sigma>0$.  So the WEC is not violated
and the collapse of the wormhole cannot be prevented
(see Figure 17).  In other words, if the exotic matter is
removed, the wormhole will collapse.
\begin{figure}
\begin{center}
\vspace{0.5cm}
\includegraphics[width=0.4\textwidth]{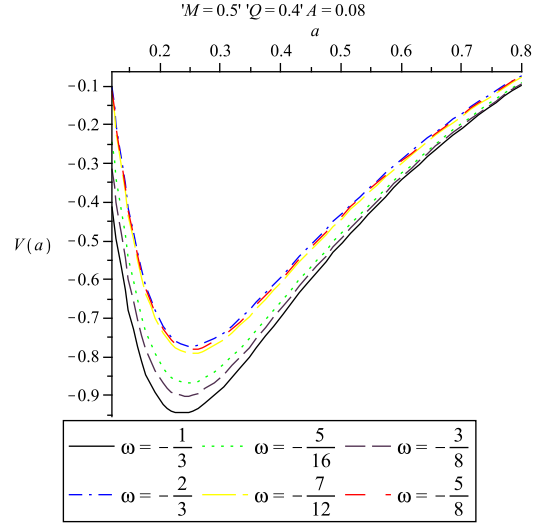}
\caption{Wormholes are stable. }
        \label{fig17}
\end{center}
\end{figure}
\begin{figure}
\begin{center}
\vspace{0.5cm}
\includegraphics[width=0.4\textwidth]{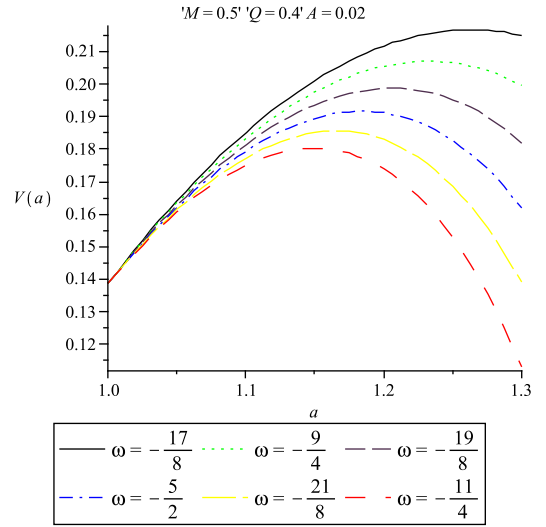}
\caption{Wormholes are unstable.}
        \label{fig17}
\end{center}
\end{figure}

\subsection{Linearized stability} \noindent
Now we will focus our attention on the stability of the
configuration under small perturbations around a static
solution at $a=a_0$.  Expanding $V(a)$ around $a_0$,
we obtain
\begin{eqnarray}
V(a) &=&  V(a_0) + V^\prime(a_0) ( a - a_0) +
\frac{1}{2} V^{\prime\prime}(a_0) ( a - a_0)^2  \nonumber \\
&\;& + O\left[( a - a_0)^3\right],
\end{eqnarray}
where the prime denotes the derivative with respect to $a$.
Since we are linearizing around $ a = a_0 $, we require that
$ V(a_0) = 0 $ and $ V^\prime(a_0)= 0 $.  The configuration
will be in stable equilibrium if $ V^{\prime\prime}(a_0)> 0 $.
The subsequent analysis will depend on a parameter $\beta$,
which is usually interpreted as the subluminal speed of sound
and is given by the relation
\[
 \beta^2(\sigma) =\left. \frac{ \partial
p}{\partial \sigma}\right\vert_\sigma.
\]
To that end, we start with Eq. (\ref{E:conservation}) and
deduce that $(a\sigma)'=-(\sigma+2p)$.  Also,
\begin{multline*}
  (a\sigma)''=-(\sigma'+2p')=-\sigma'\left(1+2
     \frac{\partial p}{\partial\sigma}\right)\\
   =2\left(1+2\frac{\partial p}{\partial\sigma}\right)
    \frac{\sigma+p}{a}=2(1+2\beta^2)\frac{\sigma+p}{a}.
\end{multline*}
Returning to Eq. (\ref{E:potential}), we now obtain
\[
  V'(a)=f'(a)+8\pi^2a\sigma(\sigma+2p)
\]
and
\begin{multline*}
   V''(a)=f''(a)-8\pi^2(\sigma+2p)^2\\
   -8\pi^2[2\sigma(1+\beta^2)(\sigma+p)].
\end{multline*}
When evaluating at the static solution $a=a_0$, we get the expected
results $V(a_0)=0$ and $V'(a_0)=0$.  The stability condition
$V''(a_0)>0$ now yields the intermediate result
\begin{equation}\label{E:intermediate}
  2\sigma(\sigma+p)(1+2\beta^2)<\frac{f''(a_0)}{8\pi^2}
    -(\sigma+2p)^2.
\end{equation}
Since both $\sigma$ and $\sigma+p$ are negative, we retain the
sense of the inequality to get
\begin{equation}\label{E:stability1}
  \beta^2<\frac{\frac{f''(a_0)}{8\pi^2}-(\sigma+2p)^2
      -2\sigma(\sigma+p)}
   {2[2\sigma(\sigma+p)]}.
\end{equation}
It follows that there is only one region of stability.
Substituting for $\sigma$ and $p$ results in
\begin{equation}\label{E:stability2}
              \beta^2 < \frac{ 1}{2(a_0f_0^\prime - 2 f_0 )}\left[
              a_0f_0^\prime- 2f_0 - a_0^2 f_0^{\prime\prime} +
              \frac{a_0^2(f_0^\prime)^2}{2f_0}\right] - 1;
                 \end{equation}
here $f_0=f(a_0)$, $f'_0=f'(a_0)$, and $f''_0=f''(a_0)$.
The final step is to substitute the expressions for $f_0$,
$f'_0$, and $f''_0$ and to graph the result for various
values of the parameters to obtain the region of stability
(below the curve in each case).  According to Figures 18 and
19, stable solutions exist for a wide range of parameters.

Returning to Figure 16 in the previous subsection, if $-1<w<0$,
then $\sigma(\sigma+p)>0$ and the inequality in
(\ref{E:intermediate}) is preserved, leading to criterion
(\ref{E:stability1}).  The parametric values $M=0.5$ and
$Q=0.4$ do indeed lie in the stability region, according to
Figure 18.

For $w<-1$ in Figure 17, $\sigma(\sigma+p)<0$, so that the sense
of the inequality in (\ref{E:intermediate}) is not preserved,
taking us out of the region of stability.

As noted earlier, since one normally interprets $\beta$ as the
speed of sound, the value should not exceed 1.  According
to Figures 18 and 19, this requirement is met in the region of
stability.  The result makes an interesting contrast to the
wormhole in Ref. \cite{Poisson1995}, constructed by using two
copies of Schwarzschild spacetime.  For that wormhole there
are two separate regions of stability that do not include the
values $0<\beta^2\le 1$, i. e., the wormhole is unstable for
all values of $a_0$ whenever $\beta$ is in this range.

\begin{figure}
\begin{center}
\vspace{0.5cm}
\includegraphics[width=0.4\textwidth]{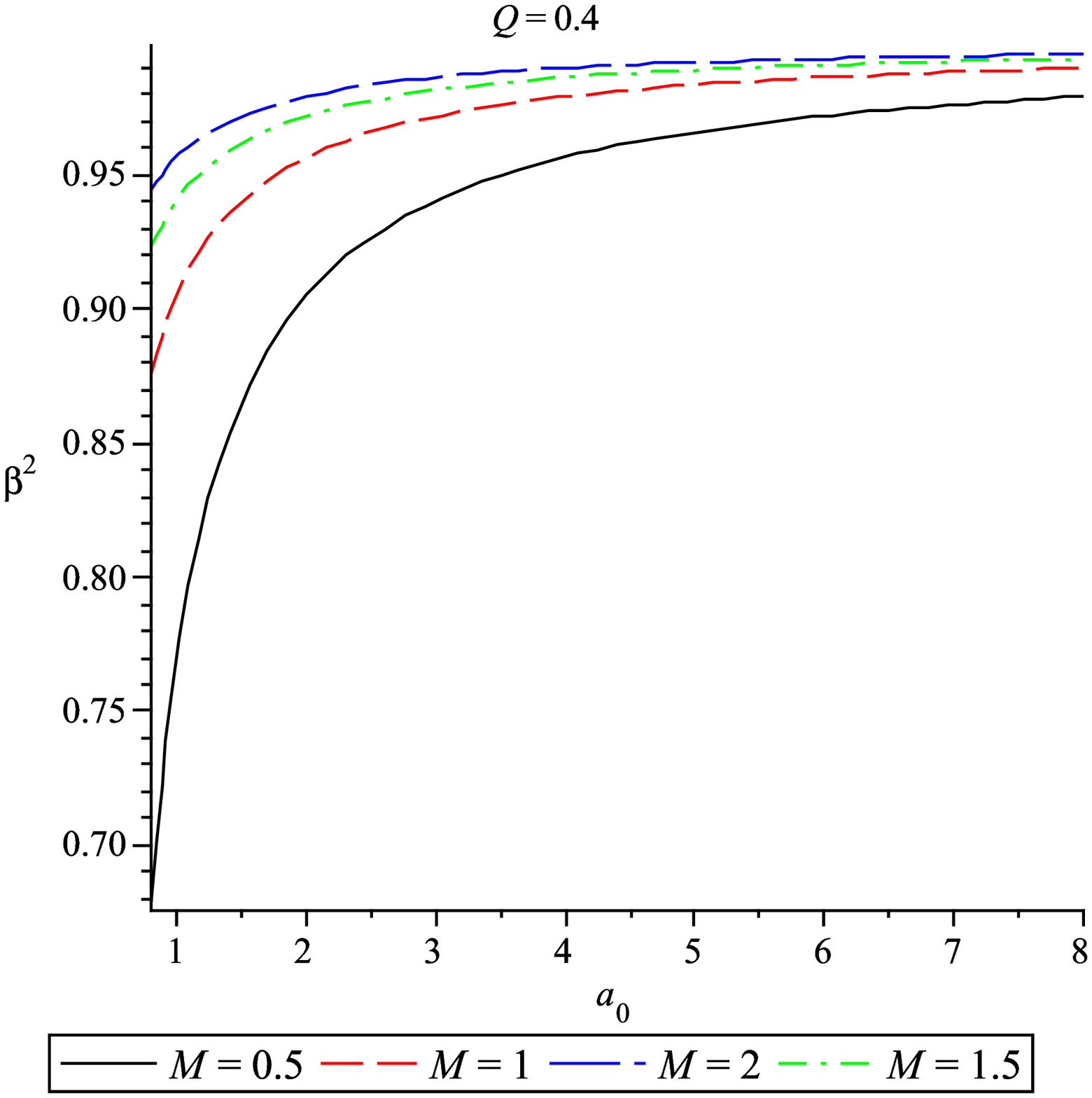}
\caption{Stability region is given  below the curve.}
        \label{fig17}
\end{center}
\end{figure}
\begin{figure}
\begin{center}
\vspace{0.5cm}
\includegraphics[width=0.4\textwidth]{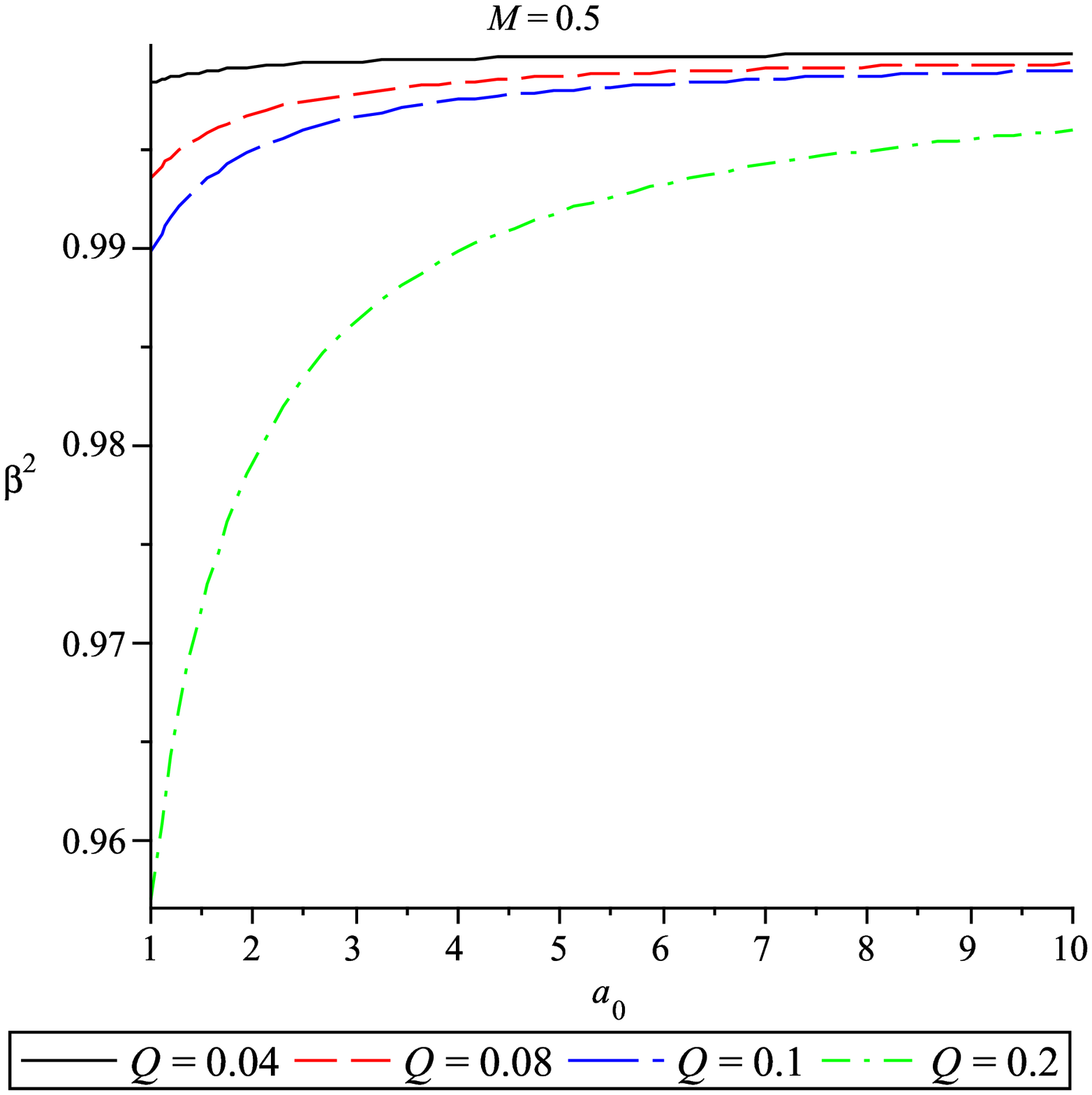}
\caption{Stability region is given  below the curve.}
        \label{fig17}
\end{center}
\end{figure}

\section{Conclusions}
\noindent This paper investigates a new thin-shell wormhole
constructed by applying the cut-and-paste technique to two
regular charged black-hole spacetimes first introduced by Beato
and Garcia.  The construction allows a graphical description of
both $\sigma$ and $p$ as functions of the radius $a$ of the thin
shell, using various values of the mass $M$ and the charge $Q$.
The same parameters help determine whether the wormhole is
attractive or repulsive.  Finally, the total amount of exotic
matter required is determined both analytically and graphically.

The issue of stability is addressed in two ways: first by
assuming the EoS $p=w\sigma$, $w<0$, on the thin shell and then
by analyzing the stability to linearized radial perturbations.
The latter yields a single region of stability covering a wide
range of values of the parameters $M, Q,$ and $a=a_0$.  The
region includes the stable solutions that depend on the EoS
$p=w\sigma$.

For our wormhole, the parameter $\beta$, which is normally interpreted as the speed of
sound, has the desired values $0<\beta^2\le 1$, unlike the wormholes
constructed from two Scharzschild spacetimes \cite{Poisson1995}.  These
are unstable for all values of $a=a_0$ in this range.


\begin{thebibliography}{99}
\bibitem{Visser1989} M. Visser, Nucl. Phys. B 328, (1989) 203.
\bibitem{Poisson1995} E. Poisson and M. Visser, Phys. Rev. D {\bf 52}, (1995) 7318.
\bibitem{Lobo2003} F.S.N. Lobo and P. Crawford, Class. Quant. Grav. {\bf 21}, (2004) 391.
\bibitem{Lobo2004} F.S.N. Lobo, Class. Quant. Grav. {\bf 21}, (2004) 4811.
\bibitem{Eiroa2004a} E.F. Eiroa and G. Romero, Gen. Rel. Grav. {\bf 36}, (2004) 651.
\bibitem{Eiroa2004b} E.F. Eiroa and C. Simeone, Phys. Rev. D {\bf 70}, (2004) 044008.
\bibitem{Eiroa2005} E.F. Eiroa and C. Simeone, Phys. Rev. D {\bf 71}, (2005) 127501.
\bibitem{Thibeault2005} M. Thibeault, C. Simeone, and E.F. Eiroa, Gen. Rel. Grav.
    {\bf 38}, (2006) 1593.
\bibitem{Lobo2005} F.S.N. Lobo, Phys. Rev. D {\bf 71}, (2005) 124022.
\bibitem{Rahaman2006} F. Rahaman et al., Gen. Rel. Grav. {\bf 38}, (2006) 1687.
\bibitem{Eiroa2007}  E.F. Eiroa and C. Simeone, Phys. Rev. D {\bf 76}, (2007) 024021.
\bibitem{Rahaman2007a} F. Rahaman et al., Int. J. Mod. Phys. D {\bf 16}, (2007) 1669.
\bibitem{Rahaman2007b} F. Rahaman et al., Gen. Rel. Grav. {\bf 39}, (2007) 945.
\bibitem{Rahaman2007c} F. Rahaman et al., Chin. J. Phys.
 {\bf45}, (2007) 518 arXiv:0705.0740 [gr-qc]
\bibitem{Richarte2007} M. G. Richarte and C. Simeone, Phys. Rev. D {\bf76},
    (2007) 087502.
\bibitem{Lemos2008} J. P. S. Lemos and F.S.N. Lobo, Phys. Rev D {\bf78},
    (2008) 044030.
\bibitem{Rahaman2008a} F. Rahaman et al., Acta Phys. Polon. B {\bf40} ,  ( 2009 ) 1575
    arXiv: gr-qc/0804.3852.
\bibitem{Rahaman2008b} F. Rahaman et al., Mod. Phys. Lett. A {\bf24},  (2009) 53 arXiv:
gr-qc/0806.1391.
\bibitem{Eiroa2008a} E.F. Eiroa, Phys. Rev. D {\bf 78}, (2008) 024018.
\bibitem{Eiroa2008b} E.F. Eiroa, M.G. Richarte, and C. Simeone, Phys. Lett. A {\bf 373}
    (2008) 1.
\bibitem{Beato1999} E. A. Beato and A. Garc\'{i}a, Phys. Lett. B  {\bf 464}, (1999) 25.
\bibitem{tR93}T.A. Roman, Phys. Rev. D {\bf 53}, (1993) 5496.
\end{thebibliography}
\end{document}